\documentclass[aip,reprint]{revtex4-1}

\draft 

\usepackage{graphicx}
\usepackage{dcolumn}
\usepackage{bm}

\begin{document}

\title{Proximity effect between a topological insulator and a magnetic insulator with large perpendicular anisotropy} 

\author{Wenmin~Yang,$^1$ Shuo~Yang,$^1$ Qinghua~Zhang,$^2$ Yang~Xu,$^1$ Shipeng~Shen,$^1$ Jian~Liao,$^1$ Jing~Teng}
\affiliation{Beijing National Laboratory for Condensed Matter Physics,
Institute of Physics, Chinese Academy of Sciences, Beijing 100190, China}
\author{Cewen~Nan}
\affiliation{Department of Materials Science and Engineering, State Key Lab of New Ceramics and Fine Processing, Tsinghua University, Beijing 100084, China}
\author{Lin~Gu,$^1$,Young~Sun}
\email{youngsun@iphy.ac.cn}
\author{Kehui~Wu}
\email{khwu@iphy.ac.cn}
\author{Yongqing~Li}
\email{yqli@iphy.ac.cn}
\affiliation
{Beijing National Laboratory for Condensed Matter Physics,
Institute of Physics, Chinese Academy of Sciences, Beijing 100190, China}
\date{\today}

\begin{abstract}
 We report that thin films of a prototype topological insulator, Bi$_{2}$Se$_{3}$, can be epitaxially grown onto the (0001) surface of BaFe$_{12}$O$_{19}$(BaM), a magnetic insulator with high Curie temperature and large perpendicular anisotropy. In the Bi$_2$Se$_3$ thin films grown on non-magnetic substrates, classic weak antilocalization (WAL) is manifested as cusp-shaped positive magnetoresistance (MR) in perpendicular magnetic fields and parabola-shaped positive MR in parallel fields, whereas in Bi$_{2}$Se$_{3}$/BaM heterostructures the low field MR is parabola-shaped, which is positive in perpendicular fields and negative in parallel fields. The magnetic field and temperature dependence of the MR is explained as a consequence of the suppression of WAL due to strong magnetic interactions at the Bi$_{2}$Se$_{3}$/BaM interface.

\end{abstract}

\maketitle 

The surface of a three-dimensional topological insulator (TI) hosts a fascinating Dirac electron system with momentum locked to real electron spins,\cite{1,2} in contrast to the valley-related pseudospins in graphene.~\cite{3} The helical spin structure has been exploited theoretically as the basis for realizing topological magnetoelectric effects and spintronic applications.~\cite{4,5,6,FuL09a,8,9,10,11,14} In many proposals, a key ingredient is to open an energy gap near the Dirac point via the proximity effect between a TI and a magnetic insulator (MI). In case of magnetization of the MI parallel to the interface, obtaining a sizable gap would require significant Fermi surface warping.~\cite{FuL09b,Oroszlany12} In contrast, an MI with out-of-plane magnetic order can break time reversal symmetry, thereby opening a large energy gap on any TI surface as long as the interfacial exchange interaction is sufficiently strong. Unfortunately, the easy magnetization axis in most known MIs, such as ferromagnets EuO,\cite{17} EuS,\cite{18,19} EuSe,\cite{20} GdN,\cite{21} and ferrimagnet yttrium iron garnet (YIG),\cite{LuYM13} lies inside the thin film/plate plane. Magnetic insulators with perpendicular magnetocrystalline anisotropy are very scarce.~\cite{23} Thus far, strong proximity effect between a TI and an MI with perpendicular anisotropy has not yet been reported, even though strong interface interaction has been realized recently in a TI/magnetically doped TI heterostructure.~\cite{FanYB14}

Here we demonstrate that Bi$_{2}$Se$_{3}$ thin films can be epitaxially grown onto BaFe$_{12}$O$_{19}$, a room temperature magnetic insulator with large perpendicular anisotropy. When a magnetic field is applied perpendicular to the Bi$_{2}$Se$_{3}$/BaFe$_{12}$O$_{19}$ heterostructure, positive magnetoresistance (MR) is observed. It has quadratic field dependence in weak magnetic fields and crosses over to logarithmic dependence in stronger fields. Applying parallel magnetic field leads to negative MR. The magnetotransport data suggest strong suppression of weak antilocalization due to the magnetic proximity effect at the Bi$_{2}$Se$_{3}$/BaFe$_{12}$O$_{19}$ interface.

M-type Barium hexaferrites (BaFe$_{12}$O$_{19}$, BaM), is an important magnetic material that has been studied for decades due to applications in magnetic recording and microwave devices.~\cite{24,25,26,27,28} It is highly insulating and has a Curie temperature of 723\,K.~\cite{28} In this work, the flat (0001) surfaces of nearly hexagon-shaped single crystalline thin plates (Fig.\,1a), were used as the substrates for epitaxial growth of Bi$_{2}$Se$_{3}$ thin films. Fig.\,1b shows magnetization curves of a typical BaM sample with magnetic field $H$ applied perpendicular and parallel to the (0001) plane at $T$=2\,K. The magnetization $M$ reaches saturation at $\mu_{0}H$=0.5\,T and 1.75\,T for perpendicular and parallel field orientations, respectively. For both orientations, $M$ has a nearly linear dependence on $H$ below the saturation. These features are in agreement with those previously reported for high quality single crystals.~\cite{24,26} The large perpendicular anisotropy, the simple $M$-$H$ relationship, and the high Curie temperature make BaM a valuable platform for investigation of the interfacial interactions between TIs and magnetic materials. Furthermore, the large remnant magnetization in some specially engineered BaM thin films~\cite{29} could be very useful for pursuing topological magnetoelectric effects without external magnetic fields.

Fig.\,1c is a high resolution cross-section transmission electron microscopy (TEM) image of a Bi$_{2}$Se$_{3}$/BaM heterostructure. It shows that the 1\,nm thick Se-Bi-Se-Bi-Se quintuple layers are parallel to the (0001) surface of BaM despite some minor ripples. The interface between BaM and Bi$_{2}$Se$_{3}$ is quite sharp, even though the first 1/2 quintuple layer is imaged less clearly than the other layers. The crystalline structure of the Bi$_{2}$Se$_{3}$/BaM heterojunction is further confirmed with x-ray diffraction, as shown in Fig.\,1d.

\begin{figure}[tbp]
\includegraphics[width=8.5cm,angle=-0]{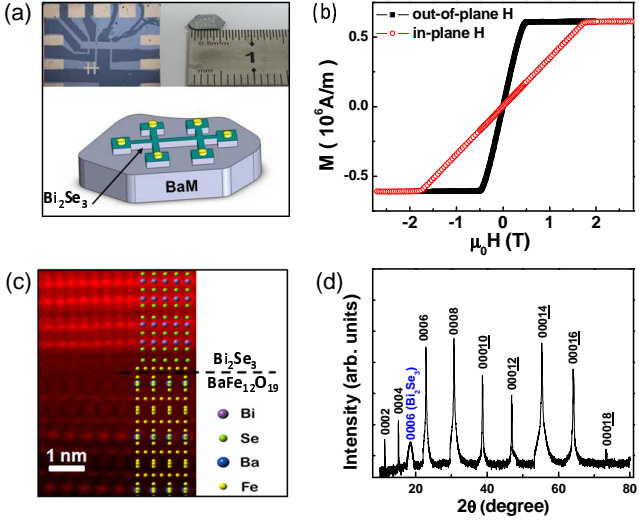}
\caption{\label{fig:epsart}(a) Schematic diagram of a Bi$_{2}$Se$_{3}$/BaM Hall bar device. The upper-left inset shows a micrograph of a 200\,$\mu$m wide Hall bar device, and the upper-right inset is an optical image of a hexagonal shaped BaM single crystalline plate with a size of  $\sim 6\times 4 \times 1$\,mm$^3$. (b) Magnetization curves measured at $T$=2\,K with an external magnetic field $H$ applied parallel (open circles) or perpendicular (solid squares) to the (0001) plane of BaM. (c) Cross-section TEM image of the interface region of a Bi$_{2}$Se$_{3}$/BaM heterostructure. (d) X-ray diffraction pattern of a Bi$_{2}$Se$_{3}$/BaM heterostructure. Diffraction peaks can be indexed either (0,0,0,2n) for BaM or (0006) for Bi$_{2}$Se$_{3}$.}
\end{figure}

Low temperature electron transport measurements were used as a probe for interfacial magnetic interactions. A thickness of 10\,nm was chosen for the Bi$_{2}$Se$_{3}$ thin films grown on the BaM substrates. Such a thickness is well above the 5\,nm threshold, below which the wavefunctions of the top and bottom surfaces overlap substantially, resulting in a hybridization gap near the Dirac point.~\cite{30} This would modify the Berry phase of the surface states, and produce transport characteristics similar to those brought by strong magnetic interactions.~\cite{31,32,32a} We also carried out transport measurements of the Bi$_{2}$Se$_{3}$ thin films grown on SrTiO$_{3}$ (STO) substrates in order to provide a reference system with non-magnetic substrates. Hall resistance $R_\mathrm{xy}$ has a nearly linear dependence on the magnetic field.~\cite{SupportInfo}
The extracted electron densities are in the range of $2$-$3\times10^{13}$\,cm$^{-2}$, consistent with previous transport and photoemission studies.~\cite{33,34,36,38,39,40,41,42} Such high electron densities indicate that the Fermi level is located above the conduction band minimum. Both the bulk and the surface electrons are expected to participate in the transport.

\begin{figure}[tbp]
\includegraphics[width=8.5cm,angle=-0]{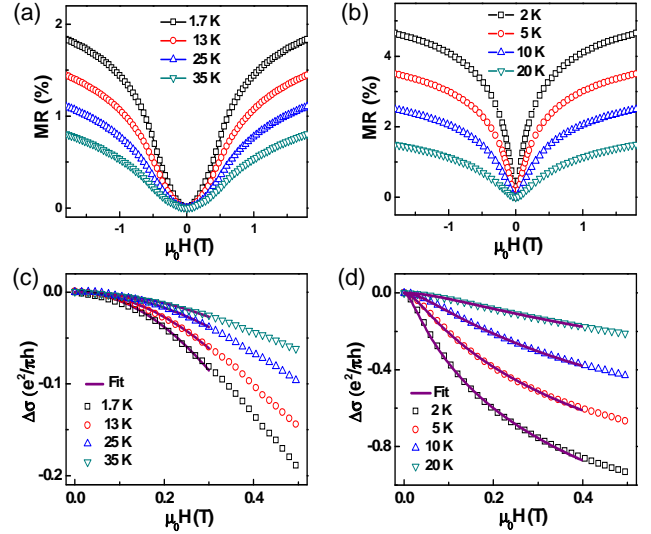}
\caption{\label{fig:epsart}Transport properties of the Bi$_{2}$Se$_{3}$ thin films grown on the BaM (a,c) and STO (b,d) substrates in perpendicular magnetic fields. (a) Magnetoresistance (MR) of a 10\,nm thick Bi$_{2}$Se$_{3}$ film on BaM at $T=1.7$-35\,K. (b) MR of a Bi$_{2}$Se$_{3}$ thin film on STO with comparable longitudinal resistivity to that of the Bi$_{2}$Se$_{3}$/BaM heterostructure. Shown in panels (c) and (d) are the corresponding magnetoconductivity data. The symbols represent experimental values. The lines in (c) and (d) are the best fits to a quadratic function and the HLN equation, respectively.}
\end{figure}

In Fig.\,2 we plot the main results of the electron transport measurements with magnetic field applied perpendicular to the Bi$_{2}$Se$_{3}$ thin films. As shown in Fig.\,2a, the magnetoresistance, defined as $MR=[\rho_\mathrm{xx}(H)-\rho_\mathrm{xx}(0)]/\rho_\mathrm{xx}(0)$, is positive for the Bi$_{2}$Se$_{3}$ thin film grown on BaM. The sign of the MR is same as that of the Bi$_{2}$Se$_{3}$ thin film on STO (Fig.\,2b). However, the shape of the MR in the Bi$_{2}$Se$_{3}$/BaM heterostructure is drastically different from its STO counterpart at low fields. The latter is characterized by the cusp-shaped MR due to the weak antilocalization (WAL) effect.\cite{34} The quantum correction to the conductivity of the Bi$_{2}$Se$_{3}$ thin films on non-magnetic substrates can be described by the Hikami-Larkin-Nagaoka (HLN) equation:~\cite{HLN}
\begin{equation}
\Delta\sigma(H)\cong-\alpha\frac{e^2}{2\pi^2\hbar}
\left[
\psi\left(\frac{1}{2}+\frac{H_\varphi}{H}\right)
-\ln\left(\frac{H_\varphi}{H}\right)
\right].
\end{equation}
Here, the magnetoconductivity is defined as $\Delta\sigma(H)=\sigma_\mathrm{xx}(H)-\sigma_\mathrm{xx}(0)$, $\psi(x)$ is the digamma function, $H_\varphi=\frac{B_\varphi}{\mu_0}=\frac{1}{\mu_{0}}\frac{\hbar}{4el_\varphi^{2}}$ is the dephasing field, and $l_{\varphi}$ is the dephasing length. In single channel systems the prefactor $\alpha$ is equal to 1/2 for WAL. As illustrated in Fig.\,2d, the $\Delta\sigma(H)$ data of the Bi$_{2}$Se$_{3}$/STO sample can be fitted fairly well to the HLN equation. The obtained $\alpha$ values are close to 1/2, which can be attributed to the strong scatterings between the surface and the bulk electrons.\cite{38,40,43} The magnetoconductivity of the Bi$_{2}$Se$_{3}$/BaM heterostructure, however, cannot be reasonably fitted to the HLN equation. As shown in Fig.\,2c, it rather exhibits a quadratic dependence on magnetic field up to at least $\mu_{0}H$=0.3\,T. At fields above the magnetization saturation of BaM (i.e.\ $\mu_{0}H>\mu_{0}H_{s}$=0.5\,T), however, the MR of the Bi$_{2}$Se$_{3}$/BaM heterostructure crosses over to the HLN-like (or logarithmic) magnetic field dependence. This implies that the phase coherent transport may still be relevant in the magnetic heterostructure.

\begin{figure}[tbp]
\includegraphics[width=8.5cm,angle=-0]{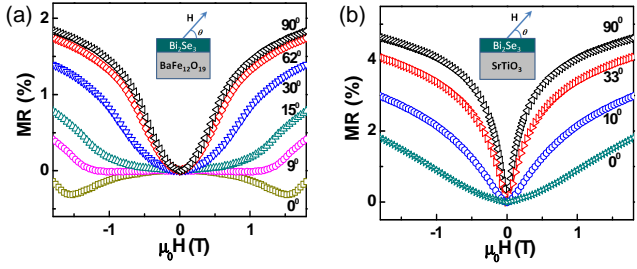}
\caption{\label{fig:epsart}Magnetoresistance (MR) data recorded in tilted magnetic fields at $T$=1.7\,K for the Bi$_{2}$Se$_{3}$/BaM heterostructure (a) and a 7\,nm thick Bi$_{2}$Se$_{3}$ thin film grown on STO (b). Data from 10\,nm thick Bi$_{2}$Se$_{3}$ samples on STO are similar, except with smaller dephasing fields.~\cite{45}}

\end{figure}

In order to gain further insight into underlying physics in the Bi$_{2}$Se$_{3}$/BaM heterostructure, we performed transport measurements in tilted magnetic fields at 1.7\,K. Fig.\,3a shows that, as the magnetic field tilts toward the thin films plane, the sign of the MR is reversed for $\theta<$9$^\mathrm{o}$, at least at $H$ below the magnetization saturation. Here, $\theta$ is the tilting angle relative to the parallel field orientation. In contrast, the MR of the Bi$_{2}$Se$_{3}$/STO remains positive for any field orientation, as shown in Fig.\,3b. In the STO case, the positive MR originates from the WAL-related phase coherent transport if the magnetic field is not too strong. Our previous work\cite{45} showed that both parallel and perpendicular components of the magnetic fileds $H$ can cause destruction of the WAL, and hence positive MR. Therefore the negative MR observed here ought to originate from the influence of the magnetic substrate.

Fig.\,4a depicts the parallel field MR of the Bi$_{2}$Se$_{3}$/BaM sample at temperatures up to 70\,K. The low field MR is negative, and has a parabolic shape. The MR reaches a minimum value at $\mu_{0}H\simeq1.55$\,T, which is close to the in-plane saturation field ($\mu_{0}H_{A}$=1.75\,T). At $H>H_{A}$, the magnetization of the BaM is aligned parallel to the interface, and one would anticipate much weaker magnetic proximity effect on the electron transport. This is evidenced by the resemblance of the parallel field MR of the Bi$_{2}$Se$_{3}$/BaM heterostructure to the STO counterpart at $H>H_{A}$ (Fig.\,4b). This further supports that the negative MR observed at lower fields is related to the magnetism in the BaM substrate.

\begin{figure}[tbp]
\includegraphics[width=8.5cm,angle=-0]{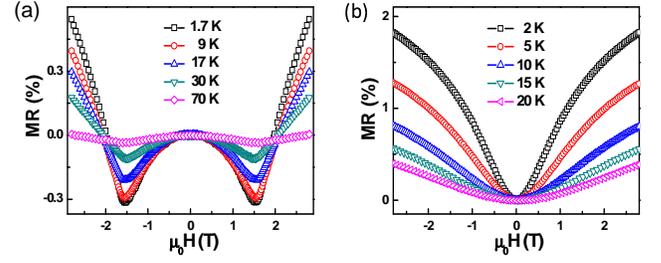}
\caption{\label{fig:epsart}The MR data taken in parallel magnetic fields. (a) MR of the Bi$_{2}$Se$_{3}$/BaM heterostructure at $T=1.7$-70\,K. (b) MR of the 7\,nm thick Bi$_{2}$Se$_{3}$ film on STO measured at $T$ up to 20\,K.}
\end{figure}

The magnetotransport data presented above can be summarized in the following two key aspects. One is the parabola-shaped MR existing in a rather broad range of magnetic fields for both parallel and perpendicular field orientations. Such a quadratic field dependence has never been observed in either perpendicular or parallel fields in previous studies of TI/MI heterostructures, such as Bi$_{2}$Se$_{3}$/EuS, Bi$_{2}$Se$_{3}$/GdN and Bi$_{2}$Se$_{3}$/YIG.~\cite{18,19,21,LangM14} The other is the strong $T$-dependence of both types of parabolic MR. This is further illustrated in Fig.\,5. For the perpendicular field orientation, the $T$-dependence of the MR is characterized by $K_{\bot}$ vs. $T$ shown in Fig.\,5a, where the coefficient $K_{\bot}$ is extracted from fitting the data in Fig.\,2c to $\Delta\sigma(B)=-K_{\bot}B^{2}$ with $B=\mu_{0}H$ up to 0.3\,T. Correspondingly, $K_{//}$ is obtained by fitting the MR data to a similar parabolic function (Fig. 5b). Both $K_{\bot}$ and $K_{//}$ drop more than three times as $T$ increases from 1.7 to 30\,K, whereas the longitudinal resistivity $\rho_\mathrm{xx}$ varies only about 10\% in the same temperature range.

As mentioned above, the MR of the Bi$_{2}$Se$_{3}$ thin films on non-magnetic substrates can be viewed as a consequence of time reversal symmetry breaking by the external perpendicular magnetic field, which introduces different Aharonov-Bohm phases to the time-reversed pairs of paths along any of the closed loops.\cite{Bergmann84} Such symmetry breaking suppresses WAL, leading to the positive, cusp-shaped MR described by the HLN equation. The parabolic MR observed in the magnetic heterostructure therefore suggests the existence of an extra source for the suppression of WAL.

In literature, random magnetic impurities\cite{HLN,Bergmann84}and magnetic exchange interaction\cite{31} are known to be able to break time reversal symmetry, and suppress the phase coherent effect. When the strength of magnetic scatterings is weak, the magnetoconductivity can also be described by the HLN equation, except that the extra dephasing due to the random magnetic scatterings needs to be taken into account.\cite{HLN,Bergmann84} The low field MR would maintain the cusp-like shape. Such behavior has been observed in GdN/Bi$_{2}$Se$_{3}$ heterostructures as well as conventional metal films (e.g. Au thin films) with magnetic adatoms.~\cite{21,Bergmann84} In case of very strong magnetic scatterings, there is a crossover from the symplectic limit ($\alpha$=1/2) to the unitary limit ($\alpha$=0).~\cite{HLN} Transport close to the latter limit~\cite{31} was observed previously in Bi$_{2}$Te$_{3}$ thin films capped with 1\,ML Fe, in which the strong magnetic scatterings from Fe nanoclusters are believed to be responsible for the parabolic MR.~\cite{48} In this classical diffusive regime, one would expect weak $T$-dependence of the MR at low temperatures. This is contradictory to the strong $T$-dependence of the MR in the Bi$_{2}$Se$_{3}$/BaM heterostructure (Fig.\,5a). Moreover, it is also unclear how the magnetic impurity scattering model can account for the negative MR in parallel fields.

For the MR in perpendicular magnetic fields, the most pronounced deviation from the WAL behavior takes place in low magnetic fields. The magnetization of the BaM substrate is featured by micron-sized maze-like domains.~\cite{24} Even though the global magnetization is small, the local magnetization has a large perpendicular component inside each domain because of the large magnetocrystalline anisotropy.~\cite{24} The magnetic exchange interaction as well as the local stray field at the interface breaks the time-reversal symmetry in the Bi$_{2}$Se$_{3}$ layer, leading to the suppression of WAL. This interface proximity effect is much larger than the conductivity corrections due to the external field. This can qualitatively explain the much weaker field dependence of the MR in the Bi$_{2}$Se$_{3}$/BaM heterostructure than that of Bi$_{2}$Se$_{3}$/STO. On a quantitative level, Lu et al.\ calculated the quantum corrections to the conductivity of TI under the influence of perpendicular magnetization. They found that in case of strong exchange interaction and weak magnetic impurity scatterings, the modified Berry phase in the surface states could result in the positive, parabolic MR.~\cite{31}

\begin{figure}[tbp]
\includegraphics[width=8.5cm,angle=-0]{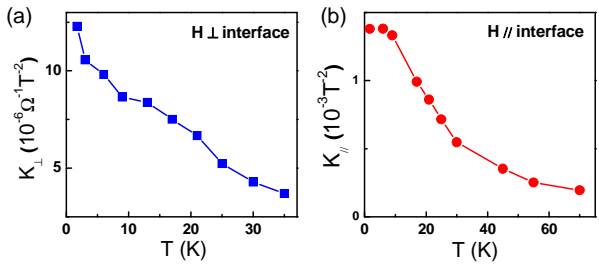}
\caption{\label{fig:epsart}Temperature dependence of the magnitude of the quadratic magnetoconductivity (or MR) in the Bi$_{2}$Se$_{3}$/BaM heterostructure in perpendicular (a) and parallel (b) magnetic fields.}
\end{figure}

The negative MR observed in parallel fields can also be qualitatively explained within the picture of broken time reversal symmetry in the phase coherent transport. Since $H$ is applied along the hard axis of BaM, it rotates the magnetization out of the perpendicular direction, and hence reduces the (local) perpendicular magnetization approximately in the form of ($1-H^{2}/H_{A}^{2}$) when $H<H_{A}$. This decreases the magnetic proximity effect, resulting in the negative MR. Nevertheless, it should be noted that the negative parabolic MR has also been observed in magnetic multilayers and magnetic granular systems, in which the MR is attributed to spin dependent scatterings.~\cite{49,50,52} In these systems, however, the MR is negative for both parallel and perpendicular field orientations. Moreover, the resistance change due to spin dependent scatterings usually has weak $T$-dependence below 30\,K. This is also inconsistent with the strong $T$-dependence of the MR of the Bi$_{2}$Se$_{3}$/BaM heterostructure (Fig.\,5). Therefore, we conclude that the MR observed in this work can be mainly attributed to the interplay between the magnetic interactions at the interface and the phase coherent transport. Nevertheless, further work is needed in order to determine whether these properties mainly originate from the interface exchange interactions or from the local stray field induced effects on the quantum diffusive transport.

In summary, we have demonstrated magnetic proximity effect in the Bi$_{2}$Se$_{3}$/BaM heterostructure. It is manifested as the parabola-shaped positive MR in perpendicular fields and negative MR in parallel fields. Such a unique type of MR has not been observed previously in any low dimensional magnetic system, including ferromagnetic thin films, magnetic multilayer structures, magnetic granular systems and TI/MI heterostructures. The proximity effect achieved in this work with the magnetic insulator that has a large perpendicular anisotropy and the Curie temperature higher than room temperature may pave a way to realizing many topological spintronic effects with potential for practical applications.

We are grateful to stimulating discussions with P. Xiong and S. von Moln\'{a}r. This work was supported by NSFC, MOST \% CAS.


\begin{thebibliography}{99}

\bibitem{1} M. Z. Hasan and C. L. Kane, Rev. Mod. Phys. \textbf{82}, 3045 (2010).
\bibitem{2} X. -L. Qi and S.-C. Zhang, Rev. Mod. Phys. \textbf{83}, 1057 (2011).
\bibitem{3} A. K. Geim and K. S. Novoselov, Nat. Mater. \textbf{6}, 183 (2007).
\bibitem{4} X. -L. Qi, T. L. Hughes, and S. -C. Zhang, Phys. Rev. B \textbf{78}, 195424 (2008).
\bibitem{5} A. M. Essin, J. E. Moore, and D. Vanderbilt,  Phys. Rev. Lett. \textbf{102}, 146805 (2009).
\bibitem{6} X.-L. Qi, R. Li, J. Zang, and S.-C. Zhang, Science \textbf{323}, 1184 (2009).
\bibitem{FuL09a} L. Fu and C. L. Kane, Phys. Rev. Lett. \textbf{102}, 216403 (2009).
\bibitem{8} A. R. Akhmerov, J. Nilsson, and C. W. J. Beenakker, Phys. Rev. Lett. \textbf{102}, 216404 (2009).
\bibitem{9} I. Garate and M. Franz, Phys. Rev. Lett. \textbf{104}, 146802 (2010).
\bibitem{10} W.-K. Tse and A. H. MacDonald, Phys. Rev. Lett. \textbf{105}, 057401 (2010).
\bibitem{11} T. Yokoyama, Y. Tanaka, and N. Nagaosa, Phys. Rev. B \textbf{81}, 121401 (2010).

\bibitem{14} G. Y. Cho and J. E. Moore, Phys. Rev. B \textbf{84}, 165101 (2011).
\bibitem{FuL09b} L. Fu, Phys. Rev. Lett. \textbf{103}, 266801 (2009).
\bibitem{Oroszlany12} L. Oroszlany and A. Cortijo, Phys. Rev. B \textbf{86}, 5427 (2012).

\bibitem{17} A. Mauger and C. Godart, Phys. Rep. \textbf{141}, 51 (1986).
\bibitem{18} P. Wei, F. Katmis, B. A. Assaf, H. Steinberg, P. Jarillo-Herrero, D. Heiman, and J. S. Moodera, Phys. Rev. Lett. \textbf{110}, 186807 (2013).
\bibitem{19} Q. I. Yang, M. Dolev, L. Zhang, J. Zhao, A. D. Fried, E. Schemm, M. Liu, A. Palevski, A. F. Marshall, S. H. Risbud, and A. Kapitulnik, Phys. Rev. B \textbf{88}, 081407 (2013).
\bibitem{20} S. von Moln\'{a}r and S. Methfess, J. Appl. Phys. \textbf{38}, 959 (1967).
\bibitem{21} A. Kandala, A. Richardella, D. W. Rench, D. M. Zhang, T. C. Flanagan, and N. Samarth, Appl. Phys. Lett. \textbf{103}, 202409 (2013).
\bibitem{LuYM13}Y. M. Lu, Y. Choi, C. M. Ortega, X. M. Cheng, J.W. Cai, S.Y. Huang,5 L. Sun, and C. L. Chien, Phys. Rev. Lett. \textbf{110}, 147207 (2013).
\bibitem{23} H. Ji, R. A. Stokes, L. D. Alegria, E. C. Blomberg, M. A. Tanatar, A. Reijnders, L. M. Schoop, T. Liang, R. Prozorov, K. S. Burch, N. P. Ong, J. R. Petta, and R. J. Cava, J. Appl. Phys. \textbf{114}, 114907 (2013).
\bibitem{FanYB14} Y. Fan, P. Upadhyaya, X. Kou, M. Lang, S. Takei, Z. Wang, J. Tang1, L. He, L.-T. Chang, M. Montazeri, G. Yu, W. Jiang, T. Nie, R. N. Schwartz, Y. Tserkovnyak, and K. L.Wang, Nat. Mater. \textbf{13}, 699 (2014).

\bibitem{24} H. Kojima and K. Goto, J. Appl. Phys. \textbf{36}, 538 (1965).
\bibitem{25} S. G. Wang, S. D. Yoon, and C. Vittoria, J. Appl. Phys. \textbf{92}, 6728 (2002).
\bibitem{26} Y. Chen, A. L. Geiler, T. Chen, T. Sakai, C. Vittoria, and V. G. Harris, J. Appl. Phys. \textbf{101}, 09M501 (2007).
\bibitem{27} Y.-Y. Song, C. L. Ordoez-Romero, and M. Wu, Appl. Phys. Lett. \textbf{95}, 142506 (2009).
\bibitem{28} Uozgur, Y. Alivov, and H. Morkoc, J. Mater. Sci.: Mater. Electron. \textbf{20}, 789 (2009).
\bibitem{29} Y. Chen, T. Sakai, T. Chen, S. D. Yoon, A. L. Geiler, C. Vittoria, and V. G. Harris, Appl. Phys. Lett. \textbf{88}, 062516 (2006).
\bibitem{30} Y. Zhang, K. He, C.-Z. Chang, C.-L. Song, L.-L. Wang, X. Chen, J.-F. Jia, Z. Fang, X. Dai, W.-Y. Shan, S.-Q. Shen, Q. Niu, X.-L. Qi, S.-C. Zhang, X.-C. Ma, and Q.-K. Xue, Nat. Phys. \textbf{6}, 584 (2010).
\bibitem{31} H.-Z. Lu, J. Shi, and S.-Q. Shen, Phys. Rev. Lett. \textbf{107}, 076801 (2011).
\bibitem{32} H.-Z. Lu, and S.-Q. Shen, Phys. Rev. B \textbf{84}, 125138 (2011).
\bibitem{32a}M. H. Liu, J. S. Zhang, C. Z. Chang, Z. C. Zhang, X. Feng, K. Li, K. He, L. L. Wang, X. Chen, X. Dai, Z. Fang, Q. K. Xue, X. C. Ma, and Y. Y. Wang, Phys. Rev. Lett. \textbf{108}, 036805 (2012).

\bibitem{SupportInfo} Supplementary material available upon request to yqli@iphy.ac.cn 

\bibitem{33} J. G. Analytis, J.-H. Chu, Y. Chen, F. Corredor, R. D. McDonald, Z. X. Shen, and I. R. Fisher, Phys. Rev. B \textbf{81}, 205407 (2010).
\bibitem{34} J. Chen, H. J. Qin, F. Yang, J. Liu, T. Guan, F. M. Qu, G. H. Zhang, J. R. Shi, X. C. Xie, C. L. Yang, K. H. Wu, Y. Q. Li, and L. Lu, Phys. Rev. Lett. \textbf{105}, 176602 (2010).

\bibitem{36} J. G. Checkelsky, Y. S. Hor, R. J. Cava, and N. P. Ong, Phys. Rev. Lett. \textbf{106}, 196801 (2011).

\bibitem{38} J. Chen, X. Y. He, K. H. Wu, Z. Q. Ji, L. Lu, J. R. Shi, J. H. Smet, and Y. Q. Li, Phys. Rev. B \textbf{83}, 241304 (2011).
\bibitem{39} Y. S. Kim, M. Brahlek, N. Bansal, E. Edrey, G. A. Kapilevich, K. Iida, M. Tanimura, Y. Horibe, S.-W. Cheong, and S. Oh, Phys. Rev. B \textbf{84}, 073109 (2011).
\bibitem{40} H. Steinberg, J. B. Lalo\"{e}, V. Fatemi, J. S. Moodera, and P. Jarillo-Herrero, Phys. Rev. B \textbf{84}, 233101 (2011).
\bibitem{41} A. A. Taskin, S. Sasaki, K. Segawa, and Y. Ando, Phys. Rev. Lett. \textbf{109}, 066803 (2012)
\bibitem{42} Y. Xia, D. Qian, D. Hsieh, L. Wray, A. Pal, H. Lin, A. Bansil, D. Grauer, Y. S. Hor, R. J. Cava, and M. Z. Hasan, Nat. Phys. \textbf{5}, 398 (2009).
\bibitem{43} I. Garate and L. Glazman, Phys. Rev. B \textbf{86}, 035422 (2012).
\bibitem{HLN} S. Hikami, A. I. Larkin, and Y. Nagaoka, Prog. Theor. Phys. \textbf{63}, 707 (1980).
\bibitem{45} C. J. Lin, X. Y. He, J. Liao, X. X. Wang, V. S. Iv, W. M. Yang, T. Guan, Q. M. Zhang, L. Gu, G. Y. Zhang, C. G. Zeng, X. Dai, K. H. Wu, and Y. Q. Li, Phys. Rev. B \textbf{88}, 041307 (2013).
\bibitem{LangM14} M. Lang, M. Montazeri, M. C. Onbasli, X. Kou, Y. Fan, P. Upadhyaya, K. Yao, F. Liu, Y. Jiang, W. Jiang, K. L. Wong, G. Yu, J. Tang, T. Nie, L. He, R. N. Schwartz, Y. Wang, C. A. Ross, and K. L. Wang, Nano Lett. \textbf{14}, 3459 (2014).
\bibitem{Bergmann84} G. Bergmann, Phys. Rep. \textbf{107}, 1 (1984).
\bibitem{48} H.-T. He, G. Wang, T. Zhang, I.-K. Sou, G. K. L. Wong, J.-N. Wang, H.-Z. Lu, S.-Q. Shen, and F.-C. Zhang, Phys. Rev. Lett. \textbf{106}, 166805 (2011).
\bibitem{49} C. L. Chien, J. Q. Xiao, and J. S. Jiang, J. Appl. Phys. \textbf{73}, 5309 (1993).
\bibitem{50} J. Q. Xiao, J. S. Jiang, and C. L. Chien, Phys. Rev. Lett. \textbf{68}, 3749 (1992).
\bibitem{52} J.-Q. Wang, P. Xiong, and G. Xiao, Phys. Rev. B \textbf{47}, 8341 (1993).

\end{thebibliography}
\end{document}